\documentclass[prl,superscriptaddress,showpacs,twocolumn]{revtex4}
\usepackage{amssymb,amsmath,amsfonts,graphicx}%,hyperref
\usepackage{hyperref}%,showkeys,ifpdf,
\begin{document}
\title{The Critical State as a Steady-State solution
of Granular Solid Hydrodynamics}
\author{Stefan Mahle}
\affiliation{Theoretische Physik, Universit\"{a}t T\"{u}bingen,
Germany}
\author{Yimin Jiang}
\affiliation{Central South University, Changsha 410083,
China}
\author{Mario Liu}
\affiliation{Theoretische Physik, Universit\"{a}t T\"{u}bingen,
Germany}
\date{\today}
\begin{abstract}
The approach to the critical state -- the transition
from partially elastic to perfectly plastic behavior
-- is considered the most characteristic of granular
phenomena in soil mechanics. By identifying the
critical state as the steady-state solution of the
elastic strain, and presenting the main results as
transparent, analytic expressions, the physics of this
important phenomenon is clarified.
\end{abstract}  \maketitle

Applying a constant shear rate to an elastic body, the
shear stress will monotonically increase -- until the
point of breakage. Granular media are different, as
they can maintain a uniform state, continually
deforming yet of constant stress -- which has, for
given density, a unique value independent of the shear
rate. This perfectly plastic state is hailed as a
hallmark of granular behavior in soil
mechanics~\cite{CSSM,PSM}, taught in introductory
courses, and called {\em critical}, to convey the idea
that it is on the verge of stability. All engineering
theories are constructed to yield a realistic
rendering of this state and the approach to it --
though the physics behind the phenomenon remains
murky. For instance, why is the {\em critical} state
at all stable, and the approach to it smooth? Or, why
is it rate-independent, but uniform dense flow is not?

In what follows, we shall first give a brief
introduction to the present thinking on the critical
state; then present granular solid hydrodynamics ({\sc
gsh}), a continual mechanical theory that, starting
from a few simple ideas about the basic physics of
sand, deduce many aspects of its behavior -- including
the critical state that is identified as a
steady-state solution of {\sc gsh}. The main ideas of
{\sc gsh} are: \textbullet~Grains with enduring
contacts are elastically deformed, which gives rise to
static shear stresses. \textbullet~The deformation is
slowly lost when grains rattle and jiggle, because
they briefly lose contact with one another. As a
consequence, a constant shear rate not only increases
the deformation, as in any elastic medium, but also
decreases it, because {grains will jiggle when sheared
past one another.} A steady state exists in which both
processes balance, such that the deformation remains
constant over time -- as does the stress. This is the
critical state and the reason for its perfect
plasticity.

The present thinking is best understood by considering
a very simple {\em elasto-plastic} model, which
patches linear elasticity with perfect plasticity:
Plotting shear stress $\sigma_s$ versus shear strain
$\varepsilon_s$, the first portion, given by
$\sigma_s\sim\varepsilon_s$, is elastic; the second,
$\sigma_s=\sigma_c=$ constant, is perfectly plastic.
As the critical stress $\sigma_c$ is equal to the
maximal elastic one, the critical state is taken to be
on the verge of stability. Consequently, the point at
which elasticity switches to plasticity is labeled the
{\em yield point}.
\begin{figure}[b]
\begin{center}
\includegraphics[scale=1.5]{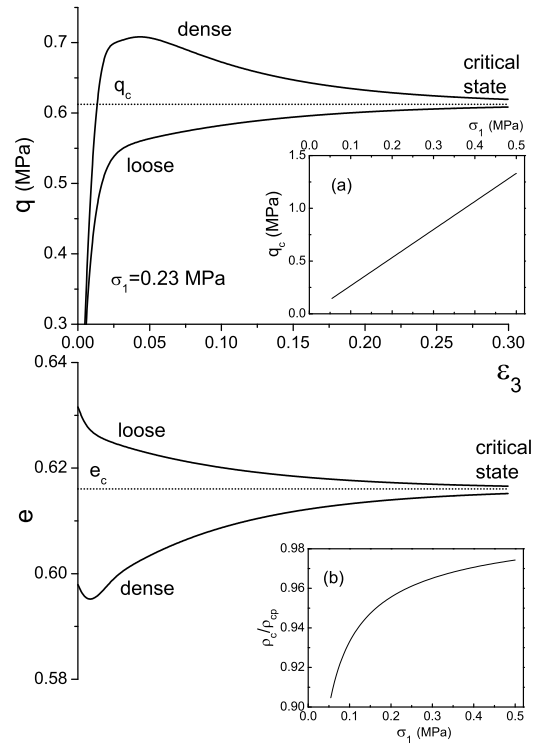}%[scale=1.34]
\end{center}
\caption{Seemingly a textbook illustration of the
critical state, this is the result of a calculation
employing {\sc gsh}, specifically the solution of
Eqs~(\ref{cs5},\ref{cs8}), plotting the stress
$q\equiv\sigma_3-\sigma_1$ and the void ratio $e$
against the strain $\varepsilon_3$ in triaxial tests
(cylinder axis along 3), at given $\sigma_1$ and
strain rate $\varepsilon_3/t$, for an initially dense
and loose sample, respectively. Insets are variations
of the critical stress $q_c$ and density $\rho_c$ with
$\sigma_1$. See the text for details, and the caption
of Fig~\ref{fig2} for the parameter values.
}\label{fig1}
\end{figure}

More realistically, one needs to include the density
dependence. Assuming cylindrical symmetry, the stress
has two independent elements, $\sigma_1=\sigma_2$ and
$\sigma_3$, in the system in which it is diagonal. The
approach to the critical state is then more varied.
The experiments are usually performed keeping one
stress element constant, say $\sigma_1$, [or
equivalently $P\equiv\frac13
(\sigma_1+\sigma_2+\sigma_3)$], while applying a
slowly increasing strain, say $\varepsilon_3$.
Employing a triaxial apparatus, one measures the
second independent stress element and the density
$\rho$. With $q\equiv\sigma_3-\sigma_1$ and the void
ratio $e\equiv\rho_g/\rho-1$ (where $\rho_g$ is the
density of the grains), the results are frequently
presented as $q(\varepsilon_3), e(\varepsilon_3)$.
Provided no shear zones are formed, the findings are
as rendered in Fig.~\ref{fig1}: (1)~$q$ increases
monotonically in loose systems, but displays a maximum
in dense ones. Both approach asymptotically the
stationary value $q_c$. (2)~Dense systems dilate,
loose systems contract, until a universal $e_c$ (or
$\rho_c$) is reached. (3)~All stress-strain curves are
rate-independent, retaining their form for whatever
strain rate. (4)~The friction angle $\varphi_c$,
defined as $\sqrt3\tan\varphi_c\equiv q_c/P$, is
essentially constant, independent of $\sigma_1$ or
$P$, implying $q_c\sim\sigma_1\sim P$. The steady
state, characterized by $q_c$ and $e_c$ for given
$\sigma_1$ or $P$, is referred to as the critical
state, see~\cite{CSSM,PSM} for more details,
and~\cite{evesq} for a presentation catered to
physicists.

There are many engineering theories capable of a
realistic account of these observations. A
mathematically elegant and comparatively simple one is
the {\em hypoplastic theory}~\cite
{Kolym-Hypoplast,PSM} that starts from a postulated
relation between the stress $\sigma_{ij}$ and strain
rate $v_{ij}\equiv\frac12(\nabla_i{v}_j
+\nabla_j{v}_i)$:
\begin{equation}\label{cs1}
{\dot\sigma }_{ij} =H_{ijk\ell}\,\,v_{k\ell}-\Lambda
_{ij}v_s,
\end{equation}
where $v_s\equiv\sqrt{v_{ij}v_{ij}}>0$. (The
compressional rate, $v_{\ell\ell}\approx0$, small in
the present context, is neglected; the dot in
$\dot\sigma_{ij}$ implies an appropriate objective
time derivative.) The tensors $H_{ijk\ell}, \Lambda
_{ij}$ are taken as functions of $\sigma_{ij},\rho$,
specified using experimental data. The critical state
is given by $\dot{\sigma }_{ij}=0$. Nevertheless,
being descriptive rather than explaining, the
hypoplastic model -- same as various elasto-plastic
ones -- is not a complete theory: As the important
features, especially rate-independence and the
so-called {\em incremental nonlinearity}, are put in
via Eq~(\ref{cs1}), there is no way to understand
them. The hypoplastic model is also narrow: None of
the rate-dependent phenomena, say dense
flow~\cite{pouliquen4}, or the transition to elastic
behavior are accounted for. (The latter is needed for
describing sound propagation~\cite{jia99} and static
stress distributions.) More fundamentally,
hypoplasticity does not specify a complete set of
state variables, and lacks energetic and entropic
considerations.

In contrast, starting from a few ideas about the
physics of sand and derived from general principles,
including energy conservation and thermodynamics, {\sc
gsh} is a full-fledged theory that aims to model
granular behavior in all its facets~\cite{GranR2}. It
was first developed to calculate static stress
distribution for various geometries, including sand
piles, silos, and point load, achieving results in
agreement with observation~\cite{ge1}. {\sc gsh} was
then generalized to dynamic situations~\cite{granL3},
producing response envelopes strikingly similar to
those from hypoplasticity. At the same time, it
clarified when rate-independence holds, and why
incremental nonlinearity applies. Two recent papers
are on elastic waves~\cite{elaWave}, showing the
quantitative agreement to experiment~\cite{jia99}, and
on {\sc gsh} presented from the point of view of soil
mechanics~\cite{GJL}. Preprints on compaction, dense
flow, fluidization and jamming have been submitted,
that on shear band is being prepared. In this Letter,
we consider the critical state.

{\sc gsh} consists of five conservation laws: for the
energy $w$, mass $\rho$, and momentum $\rho v_i$, an
evolution equation for the elastic strain $u_{ij}$,
and balance equations for two entropies, or
equivalently, two temperatures. Both the true and the
granular temperature, $T$ and $T_g$, are necessary,
because granular media sustain a {\em two-stage
irreversibility}: Macroscopic energy, kinetic and
elastic, dissipates into mesoscopic, inter-granular
degrees of freedom, which in turn degrades into
microscopic, inner-granular degrees of freedom. The
first are granular jiggling, quantified by $T_g$; the
second mostly phonons, quantified by $T$. The elastic
strain $u_{ij}$ is defined as the portion of the
strain $\varepsilon_{ij}$ that deforms the grains and
leads to reversible storage of elastic energy. The
plastic rest, $\varepsilon_{ij}-u_{ij}$, accounts for
rolling and sliding. Because the energy depends on
$u_{ij}$ alone, we take it as a state variable, while
excluding the total strain and the plastic one. This
is a crucial step that enables us to retain many
useful features of elasticity, especially an explicit
expression for the stress.

From the equations of {\sc gsh} as derived
in~\cite{GranR2}, we need three: for $T_g$, $u_{ij}$,
and the Cauchy stress $\sigma_{ij}(\rho,u_{k\ell})$.
Accounting for the first-stage irreversibility, the
balance equation for $T_g$ has a similar structure as
that for the true temperature, with the production of
granular entropy given as $R_g=\eta_gv_s^2-\gamma
T_g^2$. The first term, preceded by the viscosity
$\eta_g$, is positive. It describes how the shear rate
$v_{ij}$ jiggles grains, converting macroscopic
kinetic energy into $T_g$. The second term, negative,
accounts for inelastic collisions, as a result of
which $T_g$ diminishes with the rate $\gamma$. In
steady-state, quickly arrived at (say after $10^{-4}$
s), we have $R_g=0$, or
\begin{equation}\label{cs4}
T_g=%\sqrt{v_{ij}v_{ij}}\,\sqrt{\eta_g/\gamma}\equiv
v_s\,\sqrt{\eta_g/\gamma}.
\end{equation}
Dividing $u_{ij}$ into $\Delta\equiv-u_{\ell\ell}$ and
$u_{ij}^0\equiv
u_{ij}-\frac13u_{\ell\ell}\delta_{ij}$, the evolution
equation for $u_{ij}$ takes the
form~\cite{GranR2,footnote}
\begin{eqnarray}
\nonumber
\partial_t\Delta+(1-\alpha)v_{\ell\ell}
-\alpha_2u^0_{ij}v_{ij}
=-3\Delta/\tau_1=-3\lambda_1T_g\Delta,
\\\label{cs5}
\partial_tu^0_{ij}-(1-\alpha)v_{ij}= -u^0_{ij}/\tau
= -\lambda T_gu^0_{ij},\quad\,\,\,\,\,
\end{eqnarray}
If $T_g$ is finite, grains jiggle and briefly lose
contact with one another, during which their
unstressed form will be partially restored.
Macroscopically, this shows up as a slow relaxation of
$\Delta$ and $u_{ij}^0$, with relaxation rates that
grow with $T_g$, and  vanish for $T_g=0$, we therefore
take $1/\tau=\lambda T_g$ and $1/\tau_1=
\lambda_1T_g$. Eliminating $T_g$ using Eq~(\ref{cs4}),
we have $\lambda T_g=\Lambda v_s$,
$3\lambda_1T_g=\Lambda_1v_s$, with
$\Lambda\equiv\lambda\sqrt{{\eta_g}/{\gamma}}$,
$\Lambda_1\equiv3\lambda_1\sqrt{{\eta_g}/{\gamma}}$.
Because the relaxation of $u_{ij}$ should come to a
halt at the random closed-packed density $\rho_{cp}$,
where the system turns elastic, we take
$\Lambda,\Lambda_1\sim(\rho_{cp}-\rho)$.
$\alpha,\alpha_2>0$ are Onsager coefficients. The
first reduces the portion of $v_{ij}$ that deforms the
grains and changes the elastic strain $u_{ij}$; the
second quantifies dilatancy, the cross effect typical
of granular media that a shear flow $v_{ij}^0$ leads
to a compressional rate, $\partial_t\Delta$. For the
reason discussed below Eq~(\ref{cs10}), we also take
$\alpha_2\sim (\rho_{cp}-\rho)$. The Cauchy stress is
\begin{equation}\label{cs3}
\sigma_{ij}(u_{k\ell},\rho)=
(1-\alpha)\pi_{ij}-\alpha_2u^0_{ij}\pi_{\ell\ell}/3,
\end{equation}
where $\pi_{ij}(u_{k\ell},\rho)\equiv-\partial
w/\partial u_{ij}$, $w$ being the elastic energy.
Viscous terms (large only for the high shear rates
typical of dense flows) are neglected. That the same
$\alpha,\alpha_2$ appear here is a result of the
Onsager reciprocity relation. The term $\sim\alpha_2$
may usually be neglected, as it is an order higher in
$u_{ij}$, which is always smaller than $10^{-3}$.
%The term $\sim\alpha$ may be understood by an analogy
%to a bicycle gear shift -- a bigger $\alpha$ implies
%more pedaling at reduced torque, with the same factor.

Eqs~(\ref{cs5},\ref{cs3}) with $T_g$ eliminated are
algebraically simple and physically transparent. It is
therefore instructive that they lead directly to
Eq~(\ref{cs1}), the postulated hypoplastic relation,
and provide expressions for $H_{ijk\ell}$,
$\Lambda_{ij}$, see~\cite{GranR2}. Although these
tensors appear different from the ones used in modern
hypoplastic models, the calculated response envelopes
(ie. closed stress or strain curves) are strikingly
similar, see~\cite{granL3}. This agreement indicates
the appropriateness of our starting points, especially
the exclusion of the plastic strain as a state
variable.

Next, we evaluate the steady-state solution for the
elastic strain, given by taking $\partial_t\Delta,
\partial_tu^0_{ij}=0$ in Eqs~(\ref{cs5}). Assuming
a stationary shear flow, $v_s^2\equiv{v_{ij}v_{ij}}$,
$v_{\ell\ell}=0$, and denoting
$u_s^2\equiv{u_{ij}^0u_{ij}^0}$, we find
\begin{equation}\label{cs7}
u_c=\frac{1-\alpha}{\lambda}
\sqrt{\frac{\gamma}{\eta_g}}\equiv\frac{1-\alpha}{\Lambda}\,,\,\,\,\,
\frac{\Delta_c}{u_c}= \frac{\alpha_2}{3\lambda_1}
\sqrt{\frac{\gamma}{\eta_g}}
\equiv\frac{\alpha_2}{\Lambda_1}\,,
\end{equation}
and $u^0_{ij}/u_s=v_{ij}/v_s$. Reusing the parameter
values of~\cite{granL3}: $\Lambda=10^2$,
$\Lambda_1=30$ for $\rho/\rho_{cp}=0.96$, $\alpha=0.8$
(which are the high-$T_g$ limit of $\alpha$ and
$\eta_g/\gamma$), we obtain $u_c=2\times10^{-3}$,
${\Delta_c}/{u_c}=0.1\,\alpha_2$. As the friction
angle $\varphi_c$ is a function of ${\Delta_c}/{u_c}$
alone, see~Eq~(\ref{cs10}), it determines the value of
$\alpha_2$. The approach to  the steady state is given
by solving Eqs~(\ref{cs5}) for $u_s(t),\Delta(t)$, at
constant $\rho,v_s\equiv\varepsilon_s/t$, with the
initial conditions: $\Delta=\Delta_0, u_s=0$. The
solution is
\begin{eqnarray}\label{cs2}
u_s(t)-u_c&=&-e^{-\Lambda\varepsilon_s},\\
\Delta(t)-\Delta_c&=&\frac{\alpha_2u_c\,\,e^{-\Lambda
\varepsilon_s}}{\Lambda-\Lambda_1}+
\left[\Delta_0-\frac{\Lambda\Delta_c}{\Lambda-\Lambda_1}
\right]e^{-\Lambda_1\varepsilon_s}.\nonumber
\end{eqnarray}
As we shall soon see, these simple expressions, giving
$u_c, \Delta_c, u_s(t), \Delta(t)$ as functions of
$\rho$ and $\varepsilon_s\equiv v_st$, are a complete
account of the the critical state and the approach to
it. They describe exponential decays of $u_s(t),
\Delta(t)$ to $u_c, \Delta_c$, and are shear
rate-independent, because they depend on
$\varepsilon_s$, not $v_s$. Given $\rho,\Delta,u_s$,
the stress $\sigma_{ij}$ is known via Eq~(\ref{cs3}).

The behavior rendered in Fig~\ref{fig1} appears more
complicated, which stems from two facts of more
technical nature: First, $P$ or $\sigma_1$ is usually
held constant in stead of the density. As $\Delta,u_s$
change with time, the density compensates to maintain
$P(\rho,\Delta,u_s)$. And along with $\rho$, the
quantities $\alpha_2,\Lambda,\Lambda_1$, all functions
of $\rho$, also change with time. [In addition, there
is a change of time scale from the term
$v_{\ell\ell}=-\partial_t\rho/\rho$ in the first of
Eqs~(\ref{cs5}).]

\begin{figure}[b]
%\begin{center}
\includegraphics[scale=1.2]{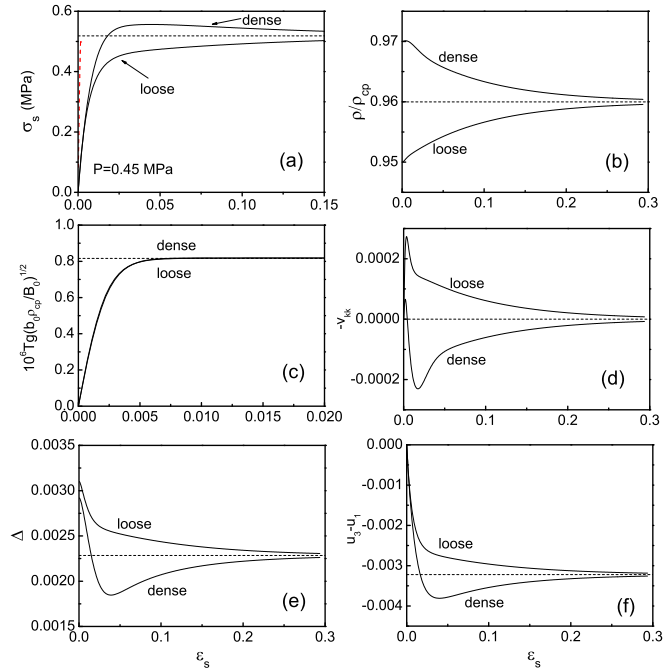}%0.6
%\end{center}
\caption{Variation of (a) the shear stress $\sigma_s$,
(b) density $\rho$, (c)~granular temperature $T_g$, (d)
volumeric deformation rate $v_{kk}$, (e) bulk strain
$\Delta$, (f) deviatoric elastic strain $u_3-u_1$, with
the total shear strain $\varepsilon_s$, of a initially
dense ($\rho_0 =0.97\rho_{cp}$) and loose ($\rho_0
=0.95\rho_{cp}$) sample, for fixed $P$ and axial
deformation rate. Initially, both $\sigma_s$ and $T_g$
vanish. Parameters: $B_0 = 7$GPa, $\xi = 5/3$,
$\rho_{\ell p}=0.85\rho_{cp}$, $\alpha_2=
750(1-\rho/\rho_{cp})$, $\alpha=0.7$,
$\Lambda=2850(1-\rho/\rho_{cp})$,
$\Lambda/\Lambda_1=3.3$, and in SI units:
$\gamma=6\times10^5T_g$, $\eta_g=0.15\,T_g$. (Note
$\Lambda=114$, $\alpha_2=30$ for $\rho =0.96\rho_{cp}$.)
The steep linear curve (red) is the purely elastic
case.}\label{fig2}
\end{figure}

Second, the stress is measured, not the elastic
strain. To calculate $\sigma_{ij}(t)$ employing
Eq~(\ref{cs3}), we need an expression for the elastic
energy $w$. The one we have consistently
employed~\cite{ge1,granL3} is: $w={\mathcal
B}\sqrt{\Delta }\left(\frac25\Delta^2+ u_s^2/\xi
\right)$ where ${\mathcal B},\xi$ are two elastic
coefficients. We fix $\xi=5/3$, independent of the
density, and take ${\cal B}(\rho)/{\cal B}_0
=\left[(\rho -\rho^*)/(\rho_{cp}-\rho)\right]^{0.15}$,
with $9\rho^*\equiv\rho_{cp}+ 20(\rho_{\ell
p}-\rho_{cp})$, and ${\cal B}_0$ around 8 GPa for
river sand, 7 GPa for glass beads. The associated
pressure $P\equiv\frac13 \sigma_{\ell\ell}$, and shear
stress $\sigma_s^2\equiv{\sigma_{ij}^0\sigma_{ij}^0}$
(with $\sigma_{ij}^0/\sigma_s=u_{ij}^0/u_s$) are
\begin{eqnarray}\label{cs8}
P&=&({1-\alpha}){\cal B}\Delta^{1.5}\left[1 +
{u_s^2}/({2\xi\Delta^2})\right],
\\\label{cs9}
{\sigma_s}&=&({1-\alpha})2u_s\sqrt\Delta\,{\cal
B}/{\xi},
\\\label{cs10}
P/{\sigma_s}&=&
{(\xi/2)\Delta}/{u_s}+{(1/4)u_s}/{\Delta},
\end{eqnarray}
where the critical pressure and shear stress are
\begin{equation}\label{cs11}
P_c=P(\rho,\Delta_c,u_c),\quad
\sigma_c=\sigma_s(\rho,\Delta_c,u_c).
\end{equation}
The expressions
Eqs~(\ref{cs7},\ref{cs2},\ref{cs8},\ref{cs9},\ref{cs11}),
evaluated for constant $\sigma_1$ or $P$, are
remarkably similar to textbook illustrations of the
critical state, see Fig~\ref{fig1} and \ref{fig2}. But
all the well-taught features of these curves are now
easy to understand: We first note that while $u_s(t)$
always increases monotonically, $\Delta(t)$ decreases
monotonically only for
$f_2\equiv[\Delta_0-\Delta_c{\Lambda}/({\Lambda-\Lambda_1})]>0$,
as $f_1\equiv\alpha_2u_c/(\Lambda-\Lambda_1)$ is
always positive. We also note that for given $P$, a
positive $f_2$ means stronger initial compression,
corresponding to a lower density or higher void ratio,
while $f_2<0$ signifies weaker compression and lower
void ratio. At the beginning, the faster relaxation of
$f_1$ dominates, so $\Delta,e$ always decrease,
irrespective of the void ratio $e$. After $f_1$ has
run its course, $\Delta, e$ go on decreasing if
$f_2>0$, but switch to increasing if $f_2<0$,
displaying respectively the so-called {\em
contractancy} and {\em dilatancy}, until the critical
state, $\Delta=\Delta_c, e=e_c$, is reached.

The shear stress $\sigma_s$ always increases first
with $u_s$, until $u_s=u_c$ is reached. The subsequent
behavior depends on what $\Delta$ does [cf.
Eq~(\ref{cs10}) and the discussion on Coulomb yield
below, consider only $u_s/\Delta<\sqrt{2\xi}$]:
$\sigma_s$ keeps growing if $\Delta$ decreases [loose
case, $f_2>0$], but decreases, displaying a peak, if
$\Delta$ grows [dense case, $f_2<0$].

The friction angle $\tan\varphi_c\equiv
q_c/\sqrt3P=\sigma_s/\sqrt2P$ is observed to be
essentially independent of the density, or the
pressure~\cite{CSSM,PSM}. This is also accounted for
by the above results, because the ratio $\sigma_s/P$
depends only on ${\Delta_c}/{u_c}=\alpha_2/\Lambda_1$,
see Eqs~(\ref{cs7}, \ref{cs10}), and we have taken
$\alpha_2,\Lambda_1\sim\rho_{cp}-\rho$. There are two
types of density dependence here: the sensitive one
via $\rho_{cp}-\rho$, and the weaker one via $\rho$.
Neglecting the latter, $\alpha_2/\Lambda_1$ is a
constant. Circumstances are similar if $P$ is the
control parameter, because $P\sim{\cal
B}\sim(\rho_{cp}-\rho)^{-0.15}$ for constant $u_{ij}$,
so changing $P$ only changes $\rho_{cp}-\rho$.

The relation of $\varphi_c$ to Coulomb yield is
instructive. Given Eq~(\ref{cs10}), the ratio
$P/\sigma_s$ has the minimal value $\sqrt{\xi/2}$, at
$u_s/\Delta=\sqrt{2\xi}$. It is larger for
$u_s/\Delta<\sqrt{2\xi}$, and unstable for
$u_s/\Delta>\sqrt{2\xi}$ (as the energy $w$ is then
concave, see~\cite{ge1}). This may be identified with
(the Druck-Prager version of the) Coulomb yield -- an
energetic instability that ensures that no elastic
solution exists for $P/\sigma_s<\sqrt{\xi/2}$, or
$u_s/\Delta>\sqrt{2\xi}$. Remarkably, this instability
is determined by the elasticity coefficient $\xi$,
while the friction angle is given by transport
coefficients, especially $\alpha_2$. There is no a
priori link between both, as they are based on
different physics. So the critical state may be stable
and observable, or not. For the first case, we have
${u_c}/{\Delta_c}=\Lambda_1/\alpha_2 <\sqrt{2\xi}$,
implying $\tan\varphi_c<1/\sqrt\xi$, or
$\varphi_c<38^\circ$ for $\xi=5/3$. (Taking as before
$\Lambda_1=30$, this is equivalent to
$\alpha_2>3\sqrt{30}$.) Clearly, if {\em critical} is
meant to imply marginal stability, it is a misnomer --
energetic stability is essential for the existence of
the critical state.

The  yield point discussed in the introduction is also
a dubious concept. Lacking an obvious choice in the
upper two curves of Fig.\ref{fig1}, one frequently
takes it as given by $q$'s maximum for dense sand --
the vague rational being the fact that in an elastic
medium, the positivity of the stiffness coefficient,
$\partial q/\partial\varepsilon_3|_{\sigma_1}=\partial
\sigma_3/\partial\varepsilon_3>0$, is required by
energetic stability, so a negative slope implies
instability and is never observed. However, since
$u_{ij}$ is the state variable and not
$\varepsilon_{ij}$, stability only requires
$\partial\sigma_3/\partial u_3$ to be positive,
freeing $\partial\sigma_3/\partial\varepsilon_3$ to be
negative, as observed.

In spite of the energetic stability of the critical
state, the yield surface may of course still be
breached at some point. Shear bands (considered in a
forthcoming paper) will then appear, destroying the
uniformity of the system. This is most likely to
happen around $q$'s maximum for the densest sand,
because $u_c$ meets the smallest $\Delta$ there.

Summary: Sheared granular media, if uniform, will
approach the continually deforming critical state that
is widely believed to be on the margin of stability.
Employing {\sc gsh}, a broadly applicable granular
theory, we find the critical state well accounted for,
with all facets transparently explained, and given by
the steady-state solution of the elastic strain that
is both continuous and  stable.

%\acknowledgements{\small\em Acknowledgement: Critical
%reading and useful remarks by Gerd Gudehus are gratefully
%acknowledged.}


\begin{thebibliography}{20}
\bibitem{footnote}
As $\alpha_2$ was neglected in~\cite{GranR2}, we revisit
the derivation. Denoting $\partial_t
u_{ij}-v_{ij}=X_{ij}$, $\sigma_{ij}=\pi_{ij}
-\sigma^D_{ij}$, the entropy production, $R=
\sigma_{ij}^Dv_{ij} + X_{ij}\pi_{ij}+\cdots$,  implies
$X_{ij}=\beta\pi_{ij}^0 +
\beta_1\delta_{ij}\pi_{\ell\ell} +\bar\alpha_{ijk\ell}
v_{k\ell}$ and $\sigma_{ij}^D=\zeta
v_{\ell\ell}\delta_{ij}+\eta
v^0_{ij}+\alpha_{ijk\ell}\pi_{k\ell}$, with the Onsager
relation, $\bar\alpha_{ijk\ell}= -\alpha_{k\ell ij}$.
Taking $\alpha_{ijk\ell}=\alpha\delta_{ik}
\delta_{j\ell}+(\alpha_2/3) u^0_{ij}\delta_{k\ell}$, ie.
including one more element than in~\cite{GranR2},
stabilizes the steady-state solution.
\end{thebibliography}
\end{document}